# Temperature-dependent Raman scattering of $KTa_{1-x}Nb_xO_3$ thin films

A. Bartasyte, J. Kreisel*

*Laboratoire Matériaux et Génie Physique, Minatec, CNRS, Grenoble Institute of Technology*
*3, parvis Louis Néel, 38016 Grenoble, France*

W. Peng, M. Guilloux-Viry

*Sciences Chimiques de Rennes, UMR 6226 CNRS / Université de Rennes 1, Campus de Beaulieu*
*35042  Rennes Cedex, France*

\* *Corresponding author:*     *J. Kreisel ([jens.kreisel@grenoble-inp.fr](jens.kreisel@grenoble-inp.fr))*

**Abstract**

We report a Raman scattering investigation of $KTa_{1-x}Nb_xO_3$ ($x$ = 0.35, 0.5) thin films deposited on MgO and $LaAlO_3$ as a function of temperature. The observed phase sequence in the range from 90 K to 720 K is similar to the structural phase transitions of the end-member material $KNbO_3$. Although similar in the phase sequence, the actual temperatures observed for phase transition temperatures are significantly different from those observed in the literature for bulk samples. Namely, the tetragonal (ferroelectric) to cubic (paraelectric) phase transition is up to 50 K higher in the films when compared to bulk samples. This enhanced ferroelectricity is attributed to biaxial strain in the investigated thin films.

The understanding of $AB$O$_3$ perovskite-type oxides is a very active research area with great relevance to both fundamental- and application-related issues, often related to their dielectric and ferroelectric properties [1,2]. In particular, ferroelectrics are well known for their potential in nonlinear optics, since their inherent non-centrosymmetric structure allows nonlinearity. Among the large family of ferroelectrics, the perovskite-type solid solution KTa$_{1-x}$Nb$_x$O$_3$ (KTN$_x$) has more recently attracted a considerable interest in the field of agile dielectric devices and for its electro-optic properties [3,4]. In an early pioneer work Triebwasser [5] has investigated the $x$-temperature phase diagram showing that KTN$_x$ undergoes for $x > 0.2$ with increasing temperature a *Rhombohedral* → *Orthorhombic* → *Tetragonal* → *Cubic* ($R → O → T → C$) phase sequence with the $R$-, $O$- and $T$-phases being ferroelectric and the $C$-phase being paraelectric. It has been established [5,6] that the paraelectric ($C$) to ferroelectric ($T$) phase transition temperature $T_c$ of bulk KTN$_x$, and thus also the related optical and dielectric properties can be tuned by changing the chemical composition.

From an application point of view, it has been realized that the synthesis of KTN$_x$ thin film is of specific interest [3,4,7,8]. However, only little is known on the effect of interfacial strain on the phase sequence and dielectric properties of KTN$_x$ films[9] even though it is well known that ferroelectric instabilities (and thus the associated properties) are easily tuned by strain[1]. Here we present an investigation of KTN$_x$ films as a function of temperature with the aim (*i*) to clarify both the room temperature (RT) structure and the temperature-dependent phase transitions of KTN$_x$ thin films with respect to bulk samples (*ii*) to investigate the effect of chemical composition on the phase sequence by comparing KTN$_{0.5}$ and KTN$_{0.35}$ thin films and (*iii*) to test the impact of a potential substrate-induced strain by comparing films deposited on MgO and LaAlO$_3$ (LAO), respectively.

To answer the above questions we have investigated KTN$_x$ thin films as a function of temperature by Raman scattering which is known to be a versatile technique for the investigation of oxide materials in particular for the detection of even subtle structural distortions in perovskites [10-14], as also recently demonstrated on similar KNa$_{1-x}$Nb$_x$O$_3$ (KNN) single crystals [15]. For the particular case of thin films, Raman scattering is known to be a powerful probe for the investigation of strain effects [16,17], texture [11], x-ray amorphous phases [12], heterostructure-related features [13,18,19] etc.

For comparison, four thin films have been investigated: KTN$_{0.5}$ and KTN$_{0.35}$ on MgO and KTN$_{0.5}$ and KTN$_{0.35}$ on LAO. All films are 300 nm thick and have been grown by pulsed laser deposition (PLD) using a KrF excimer laser. More detailed growth conditions can be found in ref. [3]. The x-ray diffraction (XRD) structural analysis showed that all samples exhibit a completely (100)-oriented perovskite phase with no detectable pyrochlore phases. Only KTN$_{0.35}$/LAO presents a very small fraction of (110) orientation. In-plane epitaxy was also verified by XRD $\phi$–scans. The out-of-plane lattice constants of films on both substrates are close to the bulk values of $KTN_{0.35}$ ($a = 3.999$ Å) and $KTN_{0.5}$ ($a = 4.003$ Å) despite the differences in lattice mismatches, i.e., a tensile strain of -4.8% on MgO and a compressive strain of 5.5% on LAO [20]. This indicates that the majority of lattice-mismatched strain is relaxed for 300 nm thick KTN films and can be ignored in residual strain. Therefore, the thermally induced strain is expected as the main source of residual strain. Note that the thermal expansion coefficients of MgO and LAO are similar ($\sim 10\times10^{-6}$ $K^{-1}$) and larger than that of KTN ($\sim 5.6\times10^{-6}$ $K^{-1}$), indicative of similar compressive strain in films. Indeed, XRD results demonstrated that the in-plane strains are similar ($\varepsilon_{//} \sim -0.3\%$) for all films except $KTN_{0.35}$/LAO sample ($\varepsilon_{//} \sim -0.10\%$). Such discrepancy for case of $KTN_{0.35}$/LAO is probably due to the formation of (110)-oriented grains which contributes an additional strain relaxation process to release more internal stress, i.e., lower residual strain in the film.

Raman spectra were recorded with a LabRam Jobin-Yvon spectrometer using the 514.5 nm lines of an $Ar^+$ ion laser. Experiments were conducted in micro-Raman with the laser focused to a 1 $\mu m^2$ spot through a times 50 long focal objective. All measurements performed under the microscope were recorded in a back-scattering geometry with the film placed in a commercial Linkam cooling/heating stage. Raman spectra were collected on the film/substrate cross section, i.e. the laser beam was parallel to the substrate plane) by using a parallel configuration of polarization (VV).

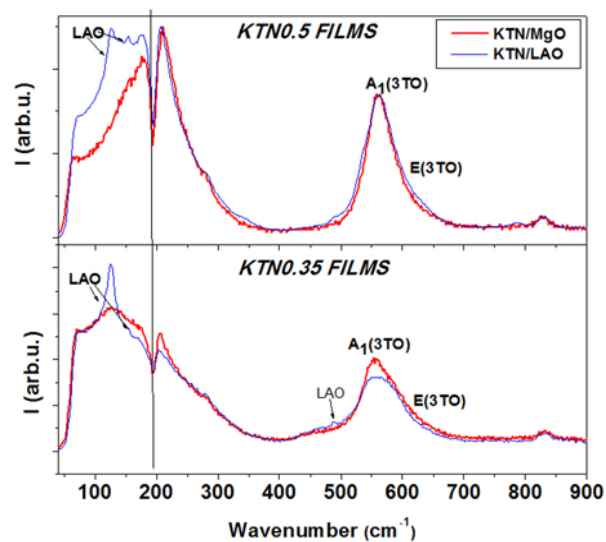

**Figure 1**

*Room temperature Raman spectra of $KTN_{0.5}$ and $KTN_{0.35}$ films on MgO and $LaAlO_3$ substrates*

**Figure 1** shows a comparison of RT spectra obtained for the four investigated $KTN_x$ films. The four spectra are at first sight rather similar with only subtle differences. The simple presence of a Raman signature indicates that the crystal structure of the four investigated $KTN_x$ films is at RT not cubic $Pm$-$3m$ for which any Raman activity is forbidden by symmetry. The Raman spectra of $KTN_x$ films at RT are similar to those obtained for the end member $KNbO_3$ (KNO) [21], although with a larger bandwidth as often observed for perovskite-type solid solutions or thin films (effects of size or defects). A comparison with spectra obtained for KNO [21], namely the presence of a resonance at around 200 cm$^{-1}$, suggests that all investigated films present a non-cubic structure at RT. According to the phase diagram of bulk KTN (ref.[5]), a tetragonal structure is indeed expected for the $KTN_{0.5}$ films. On the other hand, the structure of $KTN_{0.35}$ is expected[5] to be cubic and we attribute its observed non-cubic structure to biaxial stress resulting from the substrate-film interface, much like the strain-enhanced tetragonal phase in other perovskite films [22,23].

**Figure 2** shows the Raman spectra of the four investigated films as a function of temperature. Qualitatively, it can be seen that the Raman signature of all films is significantly modified as a function of temperature, providing direct evidence that the films undergo structural phase transitions in the investigated temperature range.

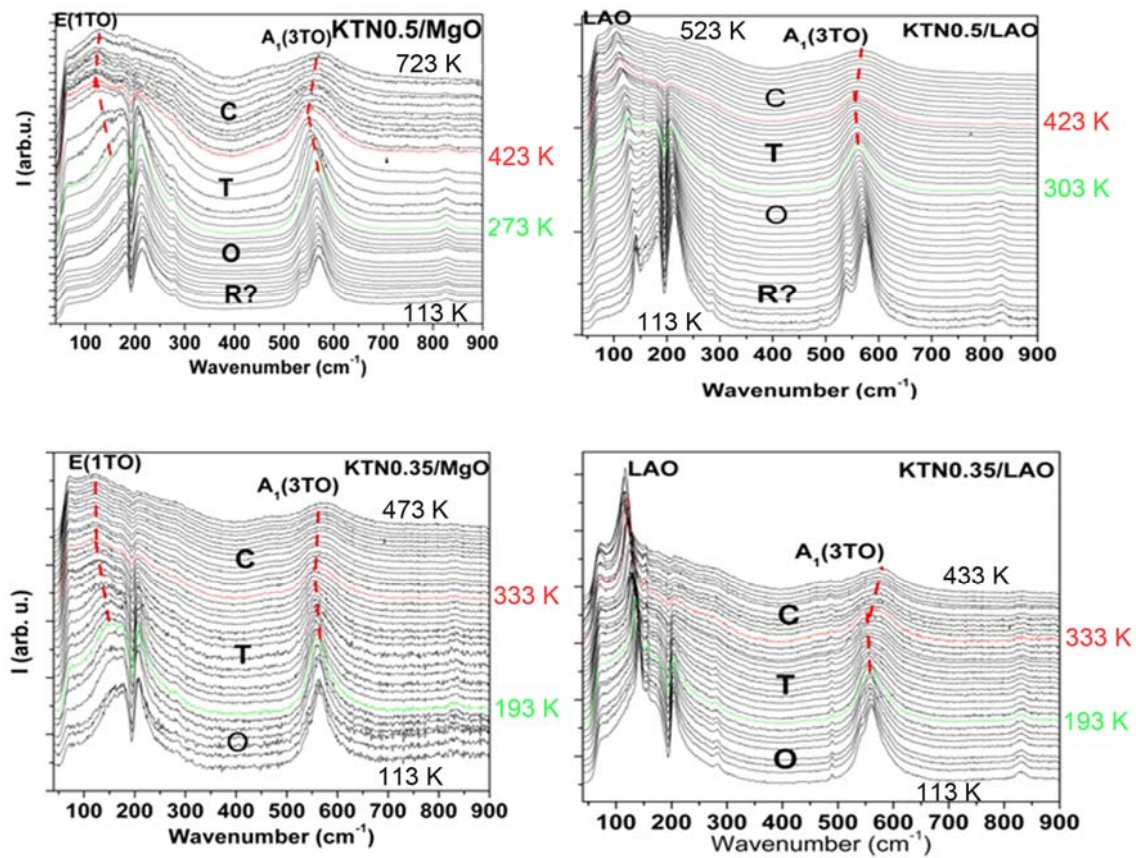

**Figure 2** (two column figure)

*Raman spectra at different temperatures at an interval of 10 K for*

*a) $KTN_{0.5}$ on MgO [top left] b) $KTN_{0.5}$ on $LaAlO_3$ [top right] c) $KTN_{0.35}$ on MgO [bottom left] and d) $KTN_{0.35}$ on $LaAlO_3$ [bottom right] substrates*

Both the RT spectrum and the temperature-dependent signature of the KTN films are qualitatively similar to that of the end member material $KNO^{21}$ but also to recent work on $(K,Na)NbO_3$ and $(K,Na)_{1-x}Li_xNbO_3$ (ref. [15]). Generally speaking, structural phase transitions in perovskites can be identified by pronounced spectral changes (soft modes, new bands, band splitting etc.) and/or rather subtle changes in wavenumber, FWHM or intensity of different bands (hard mode analysis). In the present case of KTN films we take advantage of the spectral similarity of KNO (ref. [21]) and KTN and consider the following criteria for the identification of phase transitions and their associated critical temperature $T_c$. *Rhombohedral-to-orthorhombic (R-O)*: This phase transition is difficult to be identified in our experimental setup as the expected changes in the 560 cm$^{-1}$ region are subtle and the bands cannot be resolved due to important band overlap. Thus, in the following we will not consider this phase transition for our investigated films. *Orthorhombic-to-tetragonal (O-T)*: The principal signatures for entering the tetragonal phase are the loss of the low-wavenumber wing of the 560 cm$^{-1}$ band [transformation of $B_1$(3TO) and $B_2$(3TO) to E(3TO) modes] and anomalies in the evolution of the resonance depth (see also below). Although the fit of the features around 560 cm$^{-1}$ with three overlapping bands is not straightforward, **Figure 3** illustrates for $KTN_{0.5}$/MgO that $T_c$ can however be well defined, even within a 10 K window.

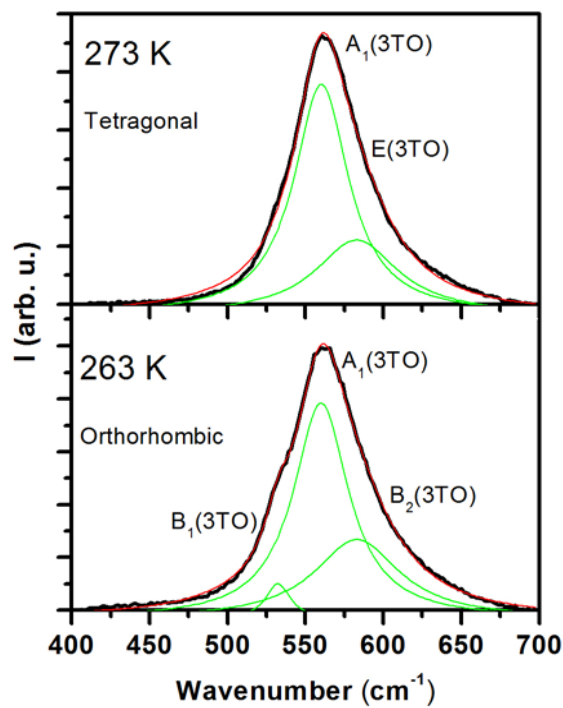

**Figure 3**

Raman spectra of KTN$_{0.5}$ on MgO at 263 K and 273 K. The spectral deconvolution (black: experiment, green: bands, red sum) illustrates the possible identification of the orthorhombic-to-tetragonal phase transition via the shoulder around 530 m$^{-1}$ (see text).

*Tetragonal-to-cubic (T-C)*: The arrival of the cubic structure is signed by the almost total disappearance of the resonance-depth around 200 cm$^{-1}$ and an anomaly in the evolution of wavenumber of the E(1TO) and A$_1$(3TO) (580 cm$^{-1}$ band) modes of the tetragonal phase (**Figure 2**). Unfortunately, in the case of KTN$_x$/LAO films, the E(1TO) mode is masked by Raman signature of LAO substrate. Further to this, the average cubic structure is characterized by a general loss in intensity by maintaining broad spectral features. Such broad features in opposition to a flat spectrum are direct evidence that the material is not strictly speaking cubic on a local level, a behaviour which is very commonly observed in ferroelectrics due to local polar fluctuations[21,24] just as in the end member KNO or BaTiO$_3$ at high-temperature.

Based on the above criteria, the phase sequence and the associated phase transition temperatures $T_c$ have been identified. **Table 1** summarises the phase assignments and $T_c$ for the different investigated samples compared to the values reported for bulk samples. The observed phase sequence is similar to the phase transitions in the end member KNO. Although the phase sequence is maintained, it can be seen that the phase diagram of the investigated films is significantly modified in terms of the phase transition temperatures when compared to bulk material. The $T_c$ of the tetragonal-to-cubic phase transition is up to 50 K higher in the films than in the bulk, which suggests that ferroelectricity (or ferroelectric instability) of the films is enhanced by biaxial stress in the thin films. This experimental observation is in line with theoretical considerations on the effect of strain on the phase transitions in epitaxial ferroelectric films: By using the Landau-Ginzburg-Devonshire type nonlinear phenomenological theory, Pertsev et al. ([25,26] and references therein) have developed the concept of a misfit strain-temperature diagram for ferroelectric films. According to this theoretical framework, the ferroelectric-paraelectric (FE-PE) phase transition temperature increases when films are subjected to tensile or compressive stresses in agreement with experimental results obtained for strained PTO/PZT films[27,28] or other perovskite-type oxide epitaxial films[22,29].

|  | $T_c$ (K) | | |
| --- | --- | --- | --- |
|  | R → O | O → T | T → C |
| KTN$_{0.35}$/LAO | ? | 190 | 330 |
| KTN$_{0.35}$/MgO | ? | 190 | 330 |
| KTN$_{0.35}$ (bulk) | 180 | 230 | 280 |
| KTN$_{0.5}$/LAO | ? | 300 | 420 |
| KTN$_{0.5}$/MgO | ? | 270 | 420 |
| KTN$_{0.5}$ (bulk) | 200 | 280 | 370 |

**Table 1**

*Phase transition temperatures deduced from Raman scattering for the different investigated KTN thin films, compared to literature results on bulk samples*

There is still an ongoing discussion in literature if KTN is a relaxor ferroelectric or not, a question which might not find the same answer for bulk or thin films[7,8,30,31]. Although Raman scattering, unlike dielectric measurements, cannot provide a definite answer as to whether a material is a relaxor or not, it is useful to remember that different types of relaxors share the characteristic that their Raman signature hardly changes with temperature: both relaxors where no long-range phase transition is observed (exemplified by PbMg$_{1/3}$Nb$_{2/3}$O$_3$, PMN [32]) and relaxors where a relaxor behaviour coexists with ferroelectric phase transitions (exemplified by PbZn$_{1/3}$Nb$_{2/3}$O$_3$, PZN [33]) show only sluggish changes. Our here observed spectral signature for KTN$_x$ thin films with changing temperature is reminiscent of Raman spectral evolutions of BaTi$_{1-x}$Zr$_x$O$_3$ (ref. [34]) in its low substitution regime where ferroelectric transitions are maintained and also to recently reported spectra of ferroelectric KNN single crystals[15]. From a Raman spectroscopy point of view this suggests (without being a formal proof) that the here investigated strained films are not relaxor ferroelectrics but rather exhibit ferroelectric phase transitions similar to pure KNbO$_3$. We note that the recent report of relaxor properties in KTa$_{0.6}$Nb$_{0.4}$O$_3$ concerns unstrained polycrystalline thin films on alumina substrates while our investigated films are at least partly strained leading to enhanced ferroelectric properties as demonstrated by the increased $T_c$ with respect to bulk samples [7].

In conclusion, we have conducted temperature-dependent Raman scattering measurements of KTa$_{1-x}$Nb$_x$O$_3$ ($x$ = 0.35, 0.5) thin films epitaxially grown on MgO and LaAlO$_3$, which provide evidence for a phase sequence similar to the structural phase transitions of the end-member material KNbO$_3$.

However, the cubic to tetragonal phase transition occurs 50 K higher than the values reported in literature for bulk. This enhanced ferroelectricity can be attributed to biaxial strain that is similar for the different samples under study.

*Acknowledgements*

Sciences Chimiques de Rennes laboratory's authors would like to thank the Region Bretagne for financial support under contract PRIR Discotec.